\documentclass[prl,preprint,showpacs]{revtex4}
\usepackage{amsmath,amsfonts}

\begin{document} 

\begin{flushright}
CP3-07-22\\
ICMPA-MPA/2007/19\\
IISc/CHEP/14/07\\
August 2007
\end{flushright}

\title{Spectrum of the non-commutative spherical well}  
\date{\today}
\author{F. G. Scholtz$^a$, B. Chakraborty$^{a,b}$, J. Govaerts$^{a,c,d,}$\footnote{Fellow of
the Stellenbosch Institute for Advanced Study (STIAS), Stellenbosch, Republic of South Africa,
{\tt http://sun.ac.za/stias}.} and S. Vaidya$^e$}
\affiliation{$^a$Institute of Theoretical Physics, University of Stellenbosch, Stellenbosch 7600, South Africa\\
$^b$S.~N.~Bose National Centre for Basic Sciences,JD Block, Sector III, Salt Lake, Kolkata-700098, India\\
$^c$Center for Particle Physics and Phenomenology (CP3),
Institut de Physique Nucl\'eaire, Universit\'e catholique de Louvain (U.C.L.),
2, Chemin du Cyclotron, B-1348 Louvain-la-Neuve, Belgium\\
$^d$UNESCO International Chair in Mathematical Physics and Applications (ICMPA),
University of Abomey-Calavi, 072 B.P. 50, Cotonou, Republic of Benin\\
$^e$Centre for High Energy Physics, Indian Institute of Science, Bangalore 560 012, India}

\begin{abstract}
We give precise meaning to piecewise constant potentials in non-commutative quantum mechanics.  In particular we discuss the infinite and finite non-commutative spherical well in two dimensions.  Using this, bound-states and scattering can be discussed unambiguously.  Here we focus on the infinite well and solve for the eigenvalues and eigenfunctions.  We find that time reversal symmetry is broken by the non-commutativity.  We show that in the commutative and thermodynamic limits the eigenstates and eigenfunctions of the commutative spherical well are recovered and time reversal symmetry is restored.  

\end{abstract}

\maketitle

There is considerable evidence coming from string theory and other approaches to the issues of quantum gravity that suggests that attempts to unify gravity and quantum mechanics will ultimately lead to a non-commutative geometry of space-time. Subsequently, non-commutative field theories and quantum mechanics have been studied extensively.  From these studies it has, however, emerged that in the presence of translational invariance and absence of interactions, non-commutativity has no physical effect \cite{gov,wess}.  However, when translational invariance is broken through the introduction of boundaries or interactions are included, the non-commutativity has physical consequences.  For example, in \cite{bem} it was shown that even in the thermodynamic limit the ground-state energy of a degenerate electron gas interacting through a screened Coulomb potential is modified to second order in perturbation theory.  This modification is entirely due to the non-commutative nature o!
 f space. In \cite{chak} it was shown for the same system that the modification in two particle correlations \cite{chak2}, arising from the twisted anti-commutation relations \cite{sachin1,sachin2}, already introduces a first order correction to the ground-state energy. The treatment of a confined gas is, however, much more difficult, both technically and conceptually.

Here we want to investigate the behaviour of an ideal non-commutative fermionic gas in a spherical well.  As translational invariance is broken, one expects non-trivial consequences due to the non-commutativity.  This is of course the simplest possible system one can study, and its commutative counterpart has played a central role in our understanding of a variety of physical phenomena, such as white dwarfs.  It is therefore natural to generalize to the non-commutative case and investigate the possible physical consequences that this may have, with the hope of experimental signatures of non-commutativity.        

The dicussion of particles moving in a non-commutative space and confined to a box or disc presents a challenge as the introduction of sharp boundaries in a non-commutative space is problematic.  Some attempts have been made in the literature to do this \cite {bal}. The most comprehensive treatment, based on our experience with the fuzzy torus and sphere, is that of  \cite{lizzia} that introduces the concept of a fuzzy disc.  Although our approach here has several elements in common with this treatment, it differs fundamentally in the way we define and solve the infinite well.  Furthermore the present treatment allows for the study of bound states in and scattering from a finite well. Our approach follows very closely the treatment of piecewise constant potentials in commutative quantum mechanics. The key element is the definition of a piecewise constant potential in the non-commutative case.  However, once that has been done, the analysis proceeds as in the commutative case!
 , one solves for the eigenstates of a constant potential and constructs the eigenstates and eigenvalues of the piecewise constant potential by matching these solutions in an appropriate way.  The derivation of these matching conditions is the key ingredient of the present treatment.  In general they are extremely complex, but it turns out that they are of a simple nature for the spherical well and that they can be treated explicitly and analytically.    

Before we can proceed, it is essential that we give precise meaning to the concepts of the classical configuration space and the Hilbert space of a non-commutative quantum system.  Elements of this can also be found in \cite{lizzia}, but we collect the essential results here for later use.  The first step is to define classical configuration space.  In two dimensions we have the non-commutative coordinate algebra 
\begin{equation}
\label{1}
[\hat x,\hat y]=i\theta,
\end{equation}
where without loss of generality it is assumed that $\theta>0$.  Using this one can introduce a pair of boson creation and annihilation operators \footnote{Following the same analysis as in the remainder of the paper, the general choice $b=[\alpha\hat{x}+i\beta\hat{y}]/\sqrt{\theta}$, $\alpha$ and $\beta$ being arbitrary complex constants such that $\alpha\beta^*+\alpha^*\beta=1$, would account for an elliptic well of which the
axes ratio and orientation in the plane are determined by the norm and phase of $\alpha/\beta$.},
$[b,b^\dagger]=1_c$,
\begin{eqnarray}
\label{2}
 b &=& \frac{1}{{\sqrt {2\theta } }}\left( {\hat x + i\hat y} \right)\nonumber \\ 
 b^\dag   &=& \frac{1}{{\sqrt {2\theta } }}\left( {\hat x - i\hat y} \right). 
\end{eqnarray}
Now we can define what we mean with classical configuration space; it is simply the boson Fock space ${\cal H}_{\rm c}  = {\rm{span}}\left\{ {\left| {n} \right\rangle } \right\}_{n = 0}^\infty$, where the span is over the field of complex numbers and, as usual, $|n\rangle=\frac{1}{\sqrt{n!}}(b^\dagger)^n|0\rangle$.  

Next we introduce the Hilbert space of the quantum non-commutative system.  This is simply defined as the set of operators ${\cal H}_{\rm q}  = \left\{ {\psi \left( {\hat x,\hat y} \right):{\rm{tr}}_c \left( {\psi \left( {\hat x,\hat y} \right)^\dag  \psi \left( {\hat x,\hat y} \right)} \right) < \infty } \right\}$. In other words, the Hilbert space is the trace class enveloping algebra of the classical configuration space Fock algebra $(b,b^\dagger)$. As these operators are necessarily bounded, this is again a Hilbert space (recall that the set of bounded operators on a Hilbert space is again a Hilbert space) and to distinguish the classical configuration space, which is also a Hilbert space, from the quantum Hilbert space we use, respectively, $c$ and $q$ as subscripts. We follow the same notation to distinguish operators acting on the classical or quantum Hilbert space. Furthermore we denote states in the quantum Hilbert space by $|\cdot)$ and states in configuration spac!
 e by $|\cdot\rangle$. The corresponding inner product is $(\psi |\phi ) = \left( {\psi ,\phi } \right){\rm{ = tr}}_{c} \left( {\psi ^\dag  \phi } \right)$, which also serves to define bra states as elements of the dual space (linear functionals).  Note that the trace is performed over the classical configuration space, denoted by subscript $c$.   
 
The next step is to construct a representation of the non-commutative Heisenberg algebra
\begin{eqnarray}
\label{heis}
 &&\left[ \hat x_q ,\hat y_q  \right] = i\theta\nonumber  \\ 
 &&\left[ \hat x_q ,\hat {p_x}_q \right] = i\hbar\nonumber  \\ 
 &&\left[ \hat y_q ,\hat {p_y}_q \right] = i\hbar  \\
 &&\left[ \hat {p_x}_q,\hat {p_y}_q \right]= 0 \nonumber
\end{eqnarray}
on the quantum Hilbert space.  This is done by defining the action of these operators as follows:
\begin{eqnarray}
\label{4}
 &&\hat x_q \psi \left( {\hat x,\hat y} \right) = \hat x\psi \left( {\hat x,\hat y} \right)\nonumber \\ 
 &&\hat y_q \psi \left( {\hat x,\hat y} \right) = \hat y\psi \left( {\hat x,\hat y} \right)\nonumber \\ 
 &&\hat {p_x}_q \psi \left( {\hat x,\hat y} \right) = \frac{\hbar }{\theta }\left[ {\hat y,\psi \left( {\hat x,\hat y} \right)} \right] \nonumber \\ 
&&\hat {p_y}_q \psi \left( {\hat x,\hat y} \right) =  - \frac{\hbar }{\theta }\left[ {\hat x,\psi \left( {\hat x,\hat y} \right)} \right] 
 \end{eqnarray}
where $\psi \left( {\hat x,\hat y} \right)$ is an arbitrary operator in the quantum Hilbert space.  Note that the momenta act adjointly.  It is easily verified, by using the Jacobi identity and non-commutative Heisenberg algebra (\ref{1}), that this is a representation.  Indeed, from the definition of the inner product it can easily be seen that this is in fact a unitary representation.  A somewhat more detailed discussion of this representation can also be found in \cite{pol}

It turns out to be more convenient to work with the complex momenta \footnote{In the case of the general elliptic well, the appropriate
definition is $p_q=\sqrt{2}[\beta\hat{p_x}_q+i\alpha\hat {p_y}_q]$.} $p_q=\hat {p_x}_q+i\hat {p_y}_q$ and $\bar p_q=\hat {p_x}_q-i\hat {p_y}_q$ ($p^2_q=\bar p_qp_q=p_q\bar p_q$),  which act as follows:
\begin{eqnarray}
\label{compmom}
 p_q \psi\left(\hat x,\hat y\right) &=& -i\hbar\sqrt{\frac{2}{\theta }}\left[b,\psi \left( {\hat x,\hat y} \right) \right], \nonumber \\ 
\bar p_q \psi\left(\hat x,\hat y\right) &=&  i\hbar\sqrt{\frac{2}{\theta }}\left[b^\dagger,\psi \left( {\hat x,\hat y} \right) \right]. 
 \end{eqnarray}

With the above notions in place one can proceed with the normal quantum mechanical interpretation in the quantum Hilbert space.  The operator $\psi(\hat x,\hat y)$ is just a vector in the quantum Hilbert space and can also be denoted $|\psi)=\psi(\hat x,\hat y)$.  A further particular useful tool in this analysis is the normalized coherent state $\left| z \right\rangle  = e^{ - \bar zz/2} e^{zb^\dag  } \left| 0 \right\rangle$, which provides an overcomplete basis on the classical configuration space. From these states one can then construct an operator, which is a vector in the quantum Hilbert space $|z)=|z\rangle \langle z|$.  The notion of a position representation, wave functions etc. can now be introduced with the conventional interpretation.  Indeed, the position representation of a state (wave function) in the quantum Hilbert space is simply $(z|\psi)={\rm tr_c}\left(|z\rangle\langle z|\psi\left(\hat x,\hat y\right)\right)=\langle z|\psi\left(\hat x,\hat y\right)|z\rangle$, with the association $z=(x+iy)/\sqrt{2\theta}$. The coherent state is also a useful technical tool for computing the inner product (traces over configuration space) explicitly.  We do not pursue this further here as the notions introduced above already suffice for our present purposes.  More detail on this can, however, be found in \cite{lizzia}.

We are interested in solving the non-commutative eigenvalue problem for a non-relativistic particle with mass $\mu$ moving in a non-commutative potential $V_q(\hat x,\hat y)$:
\begin{equation}
\label{5}
\frac{p_q^2\psi \left( {\hat x,\hat y} \right)}{2\mu}+ V_q(\hat x,\hat y)\psi \left( {\hat x,\hat y} \right)=E\psi \left( {\hat x,\hat y} \right).
\end{equation}
In particular we are interested in piecewise constant potentials, so let us first give a precise definition of this concept.  In commutative quantum mechanics we define such potentials by dividing the plane into different regions.  Introducing the characteristic functions for each region, the piecewise constant potential is simply defined as the sum of the characteristic functions of the different regions, each multiplied by the value of the potential in that region. Keeping in mind that the characteristic functions are essentially projection operators, we can easily extend this notion to the non-commutative case.  

Without loss of generality we consider two regions here, the generalization to more being obvious.  Let us therefore introduce two projection operators, $P$ and $Q$, on the classical configuration space with the properties $P^2=P$, $Q^2=Q$, $PQ=QP=0$ and $P+Q=1_c$. Here $1_c$ denotes the identity on the classical configuration space and, by definition, these operators are hermitian on the classical configuration space. Corresponding to these we define quantum operators $P_q$ and $Q_q$ with action on the quantum Hilbert space defined by $P_q\psi(\hat x,\hat y)=P\psi(\hat x,\hat y)$ and $Q_q\psi(\hat x,\hat y)=Q\psi(\hat x,\hat y)$ for any $\psi(\hat x,\hat y)$ in the quantum Hilbert space.  Note that we can also define quantum projection operators for which the projection operators act from the right.  The choice made here is dictated by the action of the potential, which is from the left, as we explain further below.  The piecewise constant potential is now simply defined as
\begin{equation}  
\label{6}
V_q=V_1P_q+V_2Q_q=P_qV_1+Q_qV_2.
\end{equation}
Here $V_1$ and $V_2$ are constants and therefore the order of writing is unimportant.   For the disc we take
\begin{equation}
\label{disc}
P=\sum_{n=0}^M |n\rangle\langle n|,\quad Q=\sum_{n=M+1}^\infty |n\rangle\langle n|.
\end{equation}  
As $\hat r^2=\hat x^2+\hat y^2=\theta(2b^\dagger b+1_c)$, $M$ determines the radius of the disc as $R^2=\theta(2M+1)$. 

Now we want to solve (\ref{5}) for this potential, which reads explicitly
\begin{equation}
\label{piecesch}
\frac{p_q^2\psi \left( {\hat x,\hat y} \right)}{2\mu}+ (V_1P_q+V_2Q_q)\psi \left( {\hat x,\hat y} \right)=E\psi \left( {\hat x,\hat y} \right).
\end{equation}
To do this let us recall the corresponding procedure in the commutative case.  There we solve the eigenvalue problem for the two constant potentials $V_1$ and $V_2$ and the same energy $E$.  Then we write a solution for the piecewise constant potential, which is the solution of the constant potential $V_1$ in region one, plus a solution of the constant potential $V_2$ in region two.  This is, however, only a solution if the two solutions and their derivatives match at the boundary.  From these matching conditions the eigenvalues of bound states, and the transmission and reflection coefficients for scattering states are then computed.

This procedure can now also be generalized quite easily to the non-commutative case.  Suppose that $\psi_1(\hat x,\hat y)$ and $\psi_2(\hat x,\hat y)$ are solutions of the constant potentials $V_1$ and $V_2$ with the same energy, {\it i.e.},
\begin{eqnarray}
\label{sch}
\frac{p_q^2\psi_1 \left( {\hat x,\hat y} \right)}{2\mu}+ V_1\psi_1 \left( {\hat x,\hat y} \right)&=&E\psi_1 \left( {\hat x,\hat y} \right),\nonumber\\
\frac{p_q^2\psi_2 \left( {\hat x,\hat y} \right)}{2\mu}+ V_2\psi_2 \left( {\hat x,\hat y} \right)&=&E\psi_2 \left( {\hat x,\hat y} \right).
\end{eqnarray}
We act from the left with $P_q$ on the first and $Q_q$ on the second, keeping in mind that the momenta do not commute with the projection operators

\begin{eqnarray}
\label{projsch}
P_q\frac{p_q^2\psi_1 \left( {\hat x,\hat y} \right)}{2\mu}+ V_1 P_q\psi_1 \left( {\hat x,\hat y} \right)&=&E P_q\psi_1 \left( {\hat x,\hat y} \right),\nonumber\\
Q_q\frac{p_q^2\psi_2 \left( {\hat x,\hat y} \right)}{2\mu}+ V_2 Q_q\psi_2 \left( {\hat x,\hat y} \right)&=&E Q_q\psi_2 \left( {\hat x,\hat y} \right).
\end{eqnarray}

Next we try to construct a solution of (\ref{piecesch}) of the form $\psi \left( {\hat x,\hat y} \right)=P_q\psi_1 \left( {\hat x,\hat y} \right)+Q_q\psi_2 \left( {\hat x,\hat y} \right)$.  The definition of the action of the quantum projection operators, namely that the projection operators act from the left, is important in this construction and dictated by the action of the potential, which is from the left in eq. (\ref{piecesch}).    Substituting this in (\ref{piecesch}), using (\ref{projsch}) and $[p_q,Q_q]=-[p_q,P_q]$, one finds that this is a solution provided that the following condition is satisfied
\begin{equation}
\label{cond}
\Omega_q \psi_1\left( {\hat x,\hat y} \right)=\Omega_q \psi_2 \left( {\hat x,\hat y} \right),
\end{equation}
where $\Omega_q$ is the operator
\begin{equation}
\label{ome}
\Omega_q=\left[p^2_q,P_q\right]=\left[p_q,P_q\right]\bar{p}_q+
\left[\bar{p}_q,P_q\right]p_q+\left[p_q,\left[\bar{p}_q,P_q\right]\right].
\end{equation}

For the disc, (see (\ref{disc})), $\Omega_q$ can be calculated easily to yield
\begin{eqnarray}
\label{omegad}
\Omega_q&=&-\frac{2\hbar^2(M+1)}{\theta}\Big(|M+1\rangle\langle M+1|-|M\rangle\langle M|\Big)+\left(i\hbar\sqrt{\frac{2\left(M+1\right)}{\theta}}|M\rangle\langle M+1|\right)\bar p_q\nonumber\\ &&+\left(i\hbar\sqrt{\frac{2\left(M+1\right)}{\theta}}|M+1\rangle\langle M|\right)p_q .
\end{eqnarray}
Using this in (\ref{cond}) and taking the inner product with $|n\rangle$, only two non-trivial conditions (when $n=M$ and $n=M+1$) survive       
\begin{eqnarray}
\label{cond1}
\left(\frac{2\hbar^2(M+1)}{\theta}\langle M|+i\hbar\sqrt{\frac{2\left(M+1\right)}{\theta}}\langle M+1|\bar p_q\right)\left(\psi_1(\hat x,\hat y)-\psi_2(\hat x,\hat y)\right)&=& 0,\nonumber\\
\left(-\frac{2\hbar^2(M+1)}{\theta}\langle M+1|+i\hbar\sqrt{\frac{2\left(M+1\right)}{\theta}}\langle M| p_q\right)\left(\psi_1(\hat x,\hat y)-\psi_2(\hat x,\hat y)\right)&= & 0.
\end{eqnarray}
This can be simplified even further.  Let us take the inner product of (\ref{cond1}) with $|\ell\rangle$ for an arbitrary $\ell$.  Using (\ref{compmom}) and the action of the creation and annihilation operators on $|\ell\rangle$ we easily arrive at 
\begin{eqnarray}
\label{mecond}
\langle M+1|\psi_1(\hat x,\hat y)|\ell+1\rangle&=&\langle M+1|\psi_2(\hat x,\hat y)|\ell+1\rangle,\quad\forall \ell\ge 0,\nonumber\\
\langle M|\psi_1(\hat x,\hat y)|\ell-1\rangle&=&\langle M|\psi_2(\hat x,\hat y)|\ell-1\rangle,\quad\forall \ell>0.
\end{eqnarray}

The only remaining task is to solve for the eigenvalues and eigenstates of the constant potential and then to apply the matching conditions (\ref{mecond}). Setting $k^2=2\mu(V-E)/\hbar^2$ the generic equation we need to solve is
\begin{equation}
\label{constsch}
p_q^2\psi \left( {\hat x,\hat y} \right)+ k^2\hbar^2\psi \left( {\hat x,\hat y} \right)=0.
\end{equation}
Here $k^2$ can be positive or negative, depending on whether one investigates bound or scattering states, respectively, and is of course different in the different domains of the piecewise constant potential.   

We start by observing that the most general form of the operator $\psi\left(\hat x,\hat y\right)$ is
\begin{equation}
\label{opeq}
\psi\left(\hat x,\hat y\right)=\sum_{k=0}^\infty\sum_{\ell=0}^\infty c_{k,\ell}(b^\dagger)^k b^\ell
\equiv\sum_{m=-\infty}^\infty\psi_m,
\end{equation}
where
\begin{eqnarray}
\label{psim}
\psi_m\equiv\sum_{k=0}^\infty c_{k,k+m}(b^\dagger)^k b^{k+m},\quad m\ge 0,&\quad&
\psi_{m}=\sum_{k=0}^\infty c_{k+|m|,k}(b^\dagger)^{k+|m|} b^{k}, \quad m<0,\nonumber\\
\left[b^\dagger b,\psi_m\right]&=&-m\psi_m,\quad\forall m.
\end{eqnarray} 
Next we observe from (\ref{compmom}) that $\left[b^\dagger b,\hat p_q^2\psi_m\right]=-m\hat p_q^2\psi_m$.  This means that operators with different values of $m$ do not get mixed by the kinetic or potential term in the constant potential Schr\"odinger equation. The solutions of (\ref{constsch}) are therefore of the form $\psi_m$, and can be labelled by $m$, which has the clear physical interpretation of angular momentum. Thus, without loss of generality we can restrict our attention to one $m$ value and consider the eigenvalue problem    
\begin{equation}
\label{constschm}
p_q^2\psi_m\left( {\hat x,\hat y} \right)+ k^2\hbar^2\psi_m \left( {\hat x,\hat y} \right)=0.
\end{equation}
Furthermore it is easy to see that $\left(p_q^2\psi\right)^\dagger=p_q^2\psi^\dagger$, which implies that if $\psi$ is a solution of (\ref{constsch}), so is $\psi^\dagger$.  From this we conclude that we can always choose the solution $\psi$ to be hermitian. This choice implies $\psi_m^\dagger=\psi_{-m}$ as can be easily seen by taking the hermitian conjugate of (\ref{opeq}) and using the linear independence of the $\psi_m$. We make this choice in what follows.

Since we need for the matching conditions (\ref{mecond}) only the matrix elements, it is sufficient if we can find explicit expressions for the matrix elements. Futhermore, since $\psi_m^\dagger=\psi_{-m}$ the matrix elements of $\psi_m$ for $m<0$ are simply related to those for $m>0$ and we only need to consider $m\ge 0$.  The only non-vanishing matrix elements of $\psi_m$ are then of the form $\langle n|\psi_m|n+m\rangle$, $n\ge 0$. Taking the matrix element of (\ref{constschm}) between the states $|n\rangle$ and $|n+m\rangle$ and using the defining properties for the action of the momenta, as well as the action of creation and annihilation operators on $|n\rangle$, we arrive at the following recursion relation for these matrix elements
\begin{eqnarray}
\label{recme1}
\left(2n+m+1+z\right)\langle n|\psi_m\left(\hat x,\hat y\right)|n+m\rangle&=&\sqrt{n(n+m)}\langle n-1|\psi_m\left(\hat x,\hat y\right)|n+m-1\rangle\nonumber\\
&+&\sqrt{(n+1)(n+m+1)}\langle n+1|\psi_m\left(\hat x,\hat y\right)|n+m+1\rangle.\nonumber\\ 
\end{eqnarray}
Here we have set $z=\frac{1}{2}\theta k^2$.  

Not surprisingly, this equation admits, apart from the trivial solution, two non-trivial independent solutions, as there are two free parameters that needs to be specified when iterating this equation.  These solutions are explicitly given by 
\begin{eqnarray}
\label{sol}
\langle n|\psi_m\left(\hat x,\hat y\right)|n+m\rangle&=&c_1(m,z)\sqrt{\frac{(n+m)!}{m!n!}}M(-n,m+1,-z),\;\forall m,n\ge 0,\nonumber\\
\langle n|\psi_m\left(\hat x,\hat y\right)|n+m\rangle&=&c_2(m,z)\sqrt{\frac{n!(n+m)!}{m!}}U(n+1,1-m,z),\;\forall m,n\ge 0,
\end{eqnarray}
where $c_1(m,z)$ and $c_2(m,z)$ are still arbitrary functions of $m$ and $z$. Furthermore $M(a,b,z)$ and $U(a,b,z)$ are the two solutions of the confluent hypergeometric differential equation and are also known as Kummer's function (\cite{abr}, p.~504). By direct substitution of the first expression into (\ref{recme1}), one easily verifies that (\ref{recme1}) is transformed into a standard recursion relation for $M(a,b,z)$ (\cite{abr}, p.~506, eq.(13.4.1)), while it is transformed into a standard recursion relation for $U(a,b,z)$ (\cite{abr}, p.~507, eq.(13.4.15)) in the case of the second expression, thus verifying that these are indeed solutions of (\ref{recme1}).  By linearity of (\ref{recme1}) we can now write the most general solution as 
\begin{eqnarray}
\label{gensol}     
&&\langle n|\psi_m\left(\hat x,\hat y\right)|n+m\rangle=\nonumber\\
&&c_1(m,z)\sqrt{\frac{m!n!}{(m+n)!}}L_n^m(-z)+c_2(m,z)\sqrt{\frac{n!(n+m)!}{m!}}U(n+1,1-m,z),\;\forall m,n\ge 0.\nonumber\\
\end{eqnarray}
Here we have also used the well known relation between Kummer's function for $a$ a negative integer and associated Laguerre polynomials (\cite{abr}, p.~509, eq.(13.6.9)).

Let us now consider the case of $\psi_m$ with $m<0$.  In this case the only non-vanishing matrix elements are of the form $\langle n-m|\psi_m|n\rangle$, $n\ge 0$.  Now $\langle n-m|\psi_m|n\rangle=\langle n|\psi^\dagger_m|n-m\rangle^*=\langle n|\psi_{-m}|n-m\rangle^*=\langle n|\psi_{|m|}|n+|m|\rangle^*$.  The corresponding matrix element for $m<0$ is therefore
\begin{eqnarray}
\label{gensoln}     
&&\langle n-m|\psi_m\left(\hat x,\hat y\right)|n\rangle=\nonumber\\
&&c_1^*(|m|,z)\sqrt{\frac{|m|!n!}{(|m|+n)!}}L_n^{|m|}(-z)+c_2^*(|m|,z)\sqrt{\frac{n!(n+|m|)!}{|m|!}}U(n+1,1-|m|,z),\nonumber\\
&&\forall m<0,n\ge 0.
\end{eqnarray}
    
Finally we have to determine the functions $c_1(m,z)$ and $c_2(m,z)$.  One way is to construct a position representation of the operator $\psi\left(\hat x,\hat y\right)$, using the coherent state basis for configuration space, and then to require the correct asymptotic behaviour at the origin and infinity.  Technically this is somewhat involved and we take an alternative, but closely related route.  We rather require that the solution  (\ref{gensol}) has the correct commutative limit.  To achieve this we note from $\hat r^2=\theta(2b^\dagger b+1)$ that $\hat r^2|n\rangle=\theta(2n+1)|n\rangle\equiv r^2|n\rangle$.  Thus the commutative limit is obtained by taking $\theta\rightarrow 0$ and $n\rightarrow\infty$, while $r$ is kept fixed.  The way we implement this limit in (\ref{gensol}) is by setting $\theta=r^2/2n$ and thus $z=r^2k^2/4n$.  Then we take the limit $n\rightarrow\infty$, while keeping $m$, the angular momentum, fixed.  

We first consider scattering states ($E>V$) and thus $k^2\equiv-\kappa^2<0$ ($\kappa>0$).  For large $n$ the asymptotic behaviour of the coefficients in (\ref{gensol}) is easily established and we have (see \cite{abr}, p.~257, eq.(6.1.39)) (we consider only $m\ge 0$ as the $m<0$ case is obtained as above)
\begin{eqnarray}
\label{gensola}     
\langle n|\psi\left(\hat x,\hat y\right)|n+m\rangle=&&c_1(m,\frac{-\kappa^2r^2}{4n})\sqrt{m!}\,n^{-m/2}L_n^m(\frac{\kappa^2r^2}{4n})+\nonumber\\
&&c_2(m,\frac{-\kappa^2r^2}{4n})\sqrt{\frac{2\pi}{m!}}e^{-n}n^{n+m/2+1/2}U(n+1,1-m,\frac{-\kappa^2r^2}{4n}).
\end{eqnarray}
Next we use the following limits (\cite{abr}, p.~787, eq.(22.15.2), p.~506, eq.(13.3.3))
\begin{eqnarray}
\label{lim}    
\mathop {\lim }\limits_{n \to \infty }n^{-m}L_n^m\left(\frac{x}{n}\right) &=&x^{-m/2}J_m\left(2\sqrt{x}\right),\nonumber\\
\mathop {\lim }\limits_{n \to \infty }\Gamma\left(n+m+1\right)U\left(n+1,1-m,\frac{x}{n+1}\right)&=&\mathop {\lim }\limits_{n \to \infty }\sqrt{2\pi}e^{-n}n^{n+m+1/2}U\left(n+1,1-m,\frac{x}{n+1}\right)\nonumber\\
&=&2x^{m/2}K_{-m}\left(2\sqrt{x}\right),
\end{eqnarray} 
where $J_m(x)$ and $K_m(x)$ are, respectively, a Bessel and modified Bessel function.  From this we conclude that we must set $c_1(m,z)=c_1(m)z^{m/2}$ and $c_2(m,z)=c_2(m)z^{-m/2}$ in order for the limit to  exist.   Then we have 
\begin{equation}
\label{lim1}    
\mathop {\lim }\limits_{n \to \infty }\langle n|\psi_m\left(\hat x,\hat y\right)|n+m\rangle=c_1(m)(-1)^{m/2}J_m\left(\kappa r\right)+\frac{c_2(m)\pi(\pm i)^{m+1}}{\sqrt{m!}}\left(J_m\left(\kappa r\right)\pm iY_m\left(\kappa r\right)\right).
\end{equation} 
Here $c_1(m)$ and $c_2(m)$ are still arbitrary $m$ dependent constants.  The $+$ sign applies when the negative root is taken, and the $-$ sign for the positive root.  Note that all the aditional constants can of course be absorbed in the arbitrary constants $c_1(m)$ and $c_2(m)$, so that they are actually irrelevant for our present discussion.  The important point to note is that inside the disc the $Y_m$ solutions are not admissable due to their singular nature at the origin.  Therefore, to conform with the correct commutative limit, we must take inside the disc $c_2(m)=0$ for all $m$. Outside the disc both solutions are admissable. This result can also be obtained by requiring the solution to be normalizable with respect to the inner product on the quantum Hilbert space.  When the trace involved in this inner product is computed using the coherent states, this translates into normalizability of the wave function.  A singularity at the origin excludes the one solution.

Next we consider bound states for which ($E<V$) so that $k^2>0$. The analysis is exactly the same as above, the only difference being that $z$ switches sign.  This leads to       
\begin{equation}
\label{lim2}    
\mathop {\lim }\limits_{n \to \infty }\langle n|\psi\left(\hat x,\hat y\right)|n+m\rangle=c_1(m)(-1)^{m/2} I_m\left(kr\right)+\frac{2c_2(m)}{\sqrt{m!}}K_m\left( kr\right).
\end{equation} 
Outside the disc the solutions $I_m$ are not admissable as they grow exponentially. Thus, to conform with the correct commutative limit we must take outside the disc $c_1(m)=0$ for all $m$. Once again this condition also follows from the normalizibility of the solution in the quantum Hilbert space. 

Now we can write down the specific solutions and matching conditions for the problem at hand.  Here we are interested in the solutions for a well with vanishing potential inside and finite potential outside, and eventually the infinite well. We first consider $m\ge 0$.  The solution inside the disc (the domain projected out by $P$ (see (\ref{disc}))) is then (recall that we have to set $c_2(m)=0$)
\begin{equation}
\label{ins}     
\langle n|\psi_{{\rm in},m}\left(\hat x,\hat y\right)|n+m\rangle=c_1(m)z_{\rm in}^{m/2}\sqrt{\frac{m!n!}{(m+n)!}}L_n^m(-z_{\rm in}),\;\forall m,n\ge 0,\,z_{\rm in}=-\frac{\mu E\theta}{\hbar^2}\equiv -\frac{\theta k_{\rm in}^2}{2}.
\end{equation}
Outside the disc we are interested in bound states with $E<V$ as we want to take the limit $V\rightarrow\infty$.  As discussed above the solutions outside for $m\ge 0$ are
\begin{eqnarray}
\label{out}     
&&\langle n|\psi_{{\rm out},m}\left(\hat x,\hat y\right)|n+m\rangle=c_2(m)z_{\rm out}^{-m/2}\sqrt{\frac{n!(n+m)!}{m!}}U(n+1,1-m,z_{\rm out}),\;\forall m,n\ge 0,\nonumber\\
&&z_{\rm out}=\frac{\mu(V-E)\theta}{\hbar^2}\equiv\frac{\theta k_{\rm out}^2}{2}.
\end{eqnarray}
Now we implement the matching conditions (\ref{mecond}), which now read (these are the only non-vanishing matrix elements) 
\begin{eqnarray}
\label{mecondm}
\langle M+1|\psi_{{\rm in},m}\left(\hat x,\hat y\right)|M+m+1\rangle&=&\langle M+1|\psi_{{\rm out},m}\left(\hat x,\hat y\right)|M+m+1\rangle,\nonumber\\
\langle M|\psi_{{\rm in},m}\left(\hat x,\hat y\right)|M+m\rangle&=&\langle M|\psi_{{\rm out},m}\left(\hat x,\hat y\right)|M+m\rangle,
\end{eqnarray}
for all $m\ge 0$.  In terms of the solutions above this reads
\begin{eqnarray}
\label{condsol}
&&c_1(m)z_{\rm in}^{m/2}\sqrt{\frac{m!(M+1)!}{(M+m+1)!}}L_{M+1}^m(-z_{\rm in})= \nonumber\\
&&c_2(m)z_{\rm out}^{-m/2}\sqrt{\frac{(M+1)!(M+m+1)!}{m!}}U(M+2,1-m,z_{\rm out}),\\
&&c_1(m)z_{\rm in}^{m/2}\sqrt{\frac{m!M!}{(m+M)!}}L_{M}^m(-z_{\rm in})=c_2(m)z_{\rm out}^{-m/2}\sqrt{\frac{M!(M+m)!}{m!}}U(M+1,1-m,z_{\rm out}).
\nonumber
\end{eqnarray}
Dividing the first condition by the second, the unknown constants $c_1(m)$ and $c_2(m)$ cancel and we find the equation for the bound state energies with positive angular momentum
\begin{equation}
\label{bsposm}
\sqrt{\frac{M+1}{M+m+1}}\frac{L_{M+1}^m\left(-z_{\rm in}\right)}{L_{M}^m\left(-z_{\rm in}\right)}\sqrt{\left(M+1\right)\left(M+m+1\right)}\frac{U\left(M+2,1-m,z_{\rm out}\right)}{U\left(M+1,1-m,z_{\rm out}\right)},\,\forall m\ge 0.
\end{equation}
The remaining free parameter, the ratio $c_2(m)/c_1(m)$, is now determined by substituting the energies back in any of the equations in (\ref{condsol}). We have not simplified eqs. (\ref{condsol}) and (\ref{bsposm}) further as the current form is convenient for discussing the commutative limit below.

One would expect that (\ref{bsposm}) would reduce to the commutative result in the commutative limit.  Let us verify that this is indeed the case.  As before the way this limit should be taken is by setting $R^2=\theta (2M+1)$.  Then the limit $M\rightarrow\infty$, $\theta\rightarrow 0$ is taken with $R^2$ held fixed.  $R>0$ then represents the radius of the commutative disc. We therefore substitute for $\theta=\frac{R^2}{2M}$, which implies the relations $z_{\rm in}=-\frac{R^2k_{\rm in}^2}{4M}$ and $z_{\rm out}=\frac{R^2k_{\rm out}^2}{4M}$ in (\ref{bsposm}).  The limit now has to be computed carefully as higher order terms need to be included.  The way to do this is to use appropriate recursive relations to express the result as ratios of Laguerre polynomials and their derivatives as well as $U$'s and their derivatives, but all of the same order. Using the recursive relations given in \cite{abr} (p.~507, eq.(13.4.23) and p.~783, eq.(22.8.6)) one easily finds that the left- !
 and right-hand sides of (\ref{bsposm}) can be expressed as
\begin{eqnarray}
\label{leftright}
{\rm LHS}&=&1+\frac{1}{M}\left(\frac{-z_{\rm in}{L^\prime}_{M+1}^m\left(-z_{\rm in}\right)}{L_{M+1}^{m}\left(-z_{\rm in}\right)}+\frac{m}{2}\right)+{\rm O}\left(\frac{1}{M^2}\right),\nonumber\\
{\rm RHS}&=&1+\frac{1}{M}\left(\frac{z_{\rm out}U^\prime\left(M+1,1-m,z\right)}{U\left(M+1,1-m,z\right)}-\frac{m}{2}\right)+{\rm O}\left(\frac{1}{M^2}\right).
\end{eqnarray}
With this result in place the limits as in (\ref{lim}) can be taken to yield the usual commutative result for a disc with radius $R$, vanishing potential on the inside and potential $V$ on the outside:
\begin{equation}
\label{commbs}
\frac{k_{\rm in}J_m^\prime\left( k_{\rm in}R\right)}{J_m\left( k_{\rm in}R\right)}=\frac{k_{\rm out}K_m^\prime
\left( k_{\rm out}R\right)}{K_m\left( k_{\rm out}R\right)}.
\end{equation}
Here $J_m$ and $K_m$ are respectively Bessel and modified Bessel functions. 

Here we are interested in the infinite well (the finite well will be discussed elsewhere \cite{sch}).  The energies of the infinite well is obtained from the limit $V\rightarrow\infty$. From the asymptotic behaviour of $U(a,b,z)$ for large $z$, (\ref{bsposm}) easily yields 
\begin{equation}
\label{infwellp}
L_{M+1}^m\left(\frac{\theta k^2}{2}\right)=0,\,\forall m\ge 0,\quad k^2=\frac{2\mu E}{\hbar^2},
\end{equation}
hence the existence of $M+1$ bound states for each positive angular momentum superselection sector, $m\ge 0$.
It is simple to see that this reduces to the usual commutative result $J_m(k R)=0$ in the commutative limit.  Note that in the infinite well the wave function vanishes outside the disc (from the asymptotics of $U(a,b,z)$ for large $z$) and that the coefficient $c_2$ is arbitrary so that one has only the energy condition (\ref{infwellp}).  This is the same as in the commutative case where the wave function also vanishes outside the disc in the infinite well limit due to the exponential damping factor.

Next we consider negative angular momenta $-m$, $m>0$.  The matching conditions are again those of (\ref{mecondm}), which follow from (\ref{mecond}), the only difference being that these matrix elements now vanish whenever $m>M$.  Let us consider the consequences of this.  The only possible non-vanishing matrix elements are $\langle M|\psi_{-m}|M-m\rangle=\langle M-m|\psi_{m}|M\rangle^*$, and similarly for $M+1$.  These matrix elements for the solution inside the well are given by
\begin{eqnarray}
\label{insn}     
\langle M|\psi_{{\rm in},-m}\left(\hat x,\hat y\right)|M-m\rangle&=&c_1^*(m)z_{\rm in}^{m/2}\sqrt{\frac{m!(M-m)!}{M!}}L_{M-m}^m(-z_{\rm in})\,\nonumber\\
\langle M+1|\psi_{{\rm in},-m}\left(\hat x,\hat y\right)|M+1-m\rangle&=&c_1^*(m)z_{\rm in}^{m/2}\sqrt{\frac{m!(M+1-m)!}{(M+1)!}}L_{M-m+1}^m(-z_{\rm in}),\nonumber\\
z_{\rm in}&=&-\frac{\mu E\theta}{\hbar^2}\equiv -\frac{\theta k_{\rm in}^2}{2}.
\end{eqnarray}
Similarly, for the solution outside they are given by          
\begin{eqnarray}
\label{outn}     
&&\langle M|\psi_{{\rm out},-m}\left(\hat x,\hat y\right)|M-m\rangle = c_2^*(m)z_{\rm out}^{-m/2}\sqrt{\frac{(M-m)!M!}{m!}}U(M-m+1,1-m,z_{\rm out}),\nonumber\\
&&\langle M+1|\psi_{{\rm out},-m}\left(\hat x,\hat y\right)|M+1-m\rangle = \nonumber \\
&&\qquad\qquad c_2^*(m)z_{\rm out}^{-m/2}\sqrt{\frac{(M+1-m)!(M+1)!}{m!}}U(M-m+2,1-m,z_{\rm out}),\nonumber\\
&&z_{\rm out} = \frac{\mu(V-E)\theta}{\hbar^2}\equiv\frac{\theta k_{\rm out}^2}{2}.
\end{eqnarray}
Now, for $m>M$ the left hand side vanishes, which implies that $c_1(m)=c_2(m)=0,\,\forall m>M$. However, if we consider the general solution (\ref{gensol}) for the non-trivial matrix elements, we note that if the coefficients $c_1$ and $c_2$ vanish for some $n$, it has to vanish for all $n$ as these coefficients are independent of $n$.  Thus all the matrix elements vanish, which can only be consistent with a trivial solution to the constant potential eigenvalue equation, {\it i.e.}, the trivial solution of the recursive relation (\ref{recme1}). Thus we conclude that solutions for the well with negative angular momentum strictly less then $-M$ have to vanish and that the angular momentum has to truncate at this point.       

The energies for the bound states with negative angular momentum $-m$, $m>0$ is now obtained as in the positive case
\begin{eqnarray}
\label{bsnegm}
&&\sqrt{\frac{M-m+1}{M+1}}\frac{L_{M-m+1}^m\left(-z_{\rm in}\right)}{L_{M-m}^m\left(-z_{\rm in}\right)}\sqrt{\left(M+1\right)\left(M-m+1\right)}\frac{U\left(M-m+2,1-m,z_{\rm out}\right)}{U\left(M-m+1,1-m,z_{\rm out}\right)},\nonumber\\
&&\forall M\ge m> 0.
\end{eqnarray}

Exactly the same analysis as for positive angular momentum shows that this also reduces to the commutative result in the commutative limit.  In the infinite well the corresponding energies for negative angular momentum $-m$, $m>0$ are given by
\begin{equation}
\label{infwelln}
L_{M-m+1}^m\left(\frac{\theta k^2}{2}\right)=0,\,\forall M\ge m> 0,\quad k^2=\frac{2\mu E}{\hbar^2},
\end{equation}
hence $1\le M+1+m\le M$ bound states in each negative angular momentum superselection sector $-M\le m\le -1$.

An immediate consequence of this result is the breaking of time reversal symmetry. Apart from the asymmetric nature of the spectrum due to the cut off in negative angular momentum, which is a direct consequence of a non vanishing non-commutative parameter $\theta$, we also observe that the energies of two states with angular momentum $\pm m$ are not the same.  There is a small splitting due to the different energy conditions (\ref{infwellp}) and (\ref{infwelln}).  It is clear that this splitting, and the asymmetry in angular momentum, dissappear in the commutative or thermodynamic limit. Indeed, we can compute this splitting to a good approximation by using the following approximation for the zeros of the Laguerre polynomials (see \cite{abr}, p.~787, eq.(22.16.8))
\begin{equation}
\label{zeroL}
L_n^m\left(x_\ell\right)=0,\quad{\rm for}\quad x_\ell=\frac{J_{m,\ell}^2}{4n+2(m+1)},\,\ell=1,2\ldots n.
\end{equation}
Here $J_{m,\ell}$ denotes the zeros of the Bessel function $J_m$.  Setting $R^2=\theta(2M+1)$ with $R>0$ the radius of the disc, we can express the energies as
\begin{eqnarray}
\label{ener}
E_\ell^m&=&\frac{E_\ell^{m,c}}{1+\frac{\theta}{R^2}(m+2)},\quad m\ge 0,\,\ell=1,2,\ldots M+1\nonumber\\
E_\ell^{-m}&=&\frac{E_\ell^{m,c}}{1+\frac{\theta}{R^2}(-m+2)},\quad M\ge m>0,\,\ell=1,2\ldots M-m+1\,
\end{eqnarray}
where $E_\ell^{m,c}$ denotes the energy of the commutative system, {\it i.e.\/}, 
\begin{equation}
\label{commen}
E_\ell^{m,c}=\frac{\hbar^2 J_{m,\ell}^2}{2\mu R^2}.
\end{equation}
Note that this energy is the same for $\pm m$.  From (\ref{ener}) it can be explicitly seen that these energies tend to the commutative ones when $\theta\rightarrow 0$ or $R\rightarrow\infty$ (the higher order terms vanish in this limit). Futhermore the splitting between $\pm m$ is easily computed as
\begin{equation}
\label{split}
\Delta E_\ell^m=E_\ell^m-E_\ell^{-m}=E_\ell^{m,c}\left(\frac{-2\frac{\theta}{R^2}m}
{\left(1+\frac{2\theta}{R^2}\right)^2-\left(\frac{\theta}{R^2}m\right)^2}\right),\,M\ge m >0.
\end{equation}
Again this vanishes in the commutative and thermodynamic limits.    

The origin of the time reveral symmetry breaking is quite clear. If one considers a non-constant hermitian potential in the Schr\"odinger equation, the term $V\psi$ is not invariant under time reversal, which now corresponds to hermitian conjugation, as $V\psi\ne \psi^\dagger V$, even though $V^\dagger=V$, which is the analogue of a real commutative potential, for which time reversal symmetry would apply in the commutative case. 

Another interesting observation from (\ref{infwellp}) and (\ref{infwelln}) is that in the extreme limit when $M=0$, and thus $R^2=\theta$, the spectrum of the infinite well is purely harmonic and only non-negative angular momentum occurs. More generally one notes from (\ref{ener}) that at low angular momenta ($m\ll M$) the energies agree to a good approximation to those of the commutative system.  Thus one would expect that when only low angular momentum states are occupied, the system will behave to a good approximation as a normal commutative system. This will be the case at low temperatures and densities. However, at large temperatures, or even more interesting high densities, the system becomes sensitive to the cut off in number of bound states per angular momentum sector, the cut off in negative angular momentum and the eigenvalues also start deviating strongly from the commutative case, so that under these conditions strong deviations from the commutative behaviour can!
  be expected.  A more detailed analysis of the thermodynamics of this system, where  these issues as well as the effects of (twisted) quantum statistics are pursued in more detail, will be presented elsewhere \cite{sch1}.        

Finally we remark that these results can also be used to study scattering from a finite well.  In this case one simply applies the matching conditions to the scattering solutions, which now have two free parameters outside the well since $c_1$ no longer needs to vanish, to compute the transmission and reflection coefficients at a given scattering energy and in a particular angular momentum channel \cite {sch}. 

We have extended the analysis of piecewise constant potentials to non-commutative systems.  The matching conditions from which the bound state energies and scattering amplitudes for a finite well can be computed have been derived.  The spectrum of the infinite well has been discussed explicitly.  The most noteworthy results are the breaking of time reversal symmetry and the restoration thereof in the commutative or thermodynamic limits.  The next important step is the generalization to higher dimensions and the study of the thermodynamics of such systems, particularly at extreme temperatures and/or densities.  Investigations in this direction are already under way.   

\noindent{\bf Acknowledgements.}
JG enjoyed a sabbatical leave at the Institute of Theoretical Physics of the
University of Stellenbosch, and is most grateful to Profs. Hendrik Geyer and Frederik Scholtz, and the School of Physics
for their warm and generous hospitality. His stay in South Africa was also supported in part by the Belgian National
Fund for Scientific Research (F.N.R.S.) through a travel grant. JG acknowledges the Abdus Salam International
Centre for Theoretical Physics (ICTP, Trieste, Italy) Visiting Scholar Programme
in support of a Visiting Professorship at the UNESCO-ICMPA (Republic of Benin).
His work is also supported by the Institut Interuniversitaire des Sciences Nucl\'eaires and by
the Belgian Federal Office for Scientific, Technical and Cultural Affairs through
the Interuniversity Attraction Poles (IAP) P6/11. FGS would like to thank Drs. Biswajit Chakraborty and Sachin Vaidya for their warm hospitality during visits at the S. N. Bose Centre and Indian Institute of Science where this work was initiated. BC would like to thank Prof. F. G. Scholtz and other members of the Institute of Theoretical Physics, Stellenbosch University for their warm hospitality during his stay there. Support under the Indo-South African research agreement between the Department of Science and Technology, Government of India and the National Research Foundation of South Africa is acknowledged, as well as a grant from the National Research Foundation of South Africa.  


\end{document}